# From Manual Android Tests to Automated and Platform Independent Test Scripts

Mattia Fazzini*, Eduardo Noronha de A. Freitas†, Shauvik Roy Choudhary*, Alessandro Orso*

*Georgia Institute of Technology
Atlanta, GA, USA
{mfazzini | shauvik | orso}@cc.gatech.edu

†Instituto Federal de Goiás
Goiânia, GO, Brazil
efreitas@ifg.edu.br

## ABSTRACT

Because Mobile apps are extremely popular and often mission critical nowadays, companies invest a great deal of resources in testing the apps they provide to their customers. Testing is particularly important for Android apps, which must run on a multitude of devices and operating system versions. Unfortunately, as we confirmed in many interviews with quality assurance professionals, app testing is today a very human intensive, and therefore tedious and error prone, activity. To address this problem, and better support testing of Android apps, we propose a new technique that allows testers to easily create platform independent test scripts for an app and automatically run the generated test scripts on multiple devices and operating system versions. The technique does so without modifying the app under test or the runtime system, by (1) intercepting the interactions of the tester with the app and (2) providing the tester with an intuitive way to specify expected results that it then encode as test oracles. We implemented our technique in a tool named BARISTA and used the tool to evaluate the practical usefulness and applicability of our approach. Our results show that BARISTA can faithfully encode user defined test cases as test scripts with built-in oracles, generates test scripts that can run on multiple platforms, and can outperform a state-of-the-art tool with similar functionality. BARISTA and our experimental infrastructure are publicly available.

## 1. INTRODUCTION

Mobile platforms are becoming increasingly prevalent, and so are the mobile applications (or simply apps) that run on such platforms. Today, we use apps for many of our daily activities, such as shopping, banking, social networking, and traveling. Like all other software applications, apps must be tested to gain confidence that they behave correctly under different inputs and conditions. This is especially important nowadays, given the number of companies that make apps available to their users, as failures in an app can result in loss of reputation, and ultimately customers, for the company that provides the app. For this reason, companies are spending considerable amounts of money and resources on quality assurance (QA) activities, and in particular on testing.

In the case of Android apps, the picture is further complicated by the fragmentation of the Android ecosystem [33], which includes countless devices that come in all shapes and sizes and that can run a number of different versions of the Android operating system. Gaining confidence that an app works correctly across the whole range of Android devices and operating system versions is especially challenging and expensive.

In the last several months, in the context of a customer-discovery exercise, we conducted a large number of interviews with QA professionals in a vast range of companies, going from startups to large corporations. Through these interviews, we discovered that there are two main ways in which mobile apps are tested nowadays. The first way is to follow a previously written test script that describes which actions to (manually) perform on the app (*e.g.,* entering a value in a text entry or pushing a button) and which results to expect in return (*e.g.,* a confirmation message after submitting some information). Writing, and especially performing, such scripts is extremely tedious, time consuming, and error prone. The second way is to encode actions and expected results in some testing framework, such as Google's Espresso [15], which is part of the Android Testing Support Library and it is becoming a de-facto standard in the Android testing world. The use of a framework can alleviate or eliminate some of the issues associated with a purely manual approach. For instance, it allows for automatically rerun test cases, possibly on multiple platforms, after encoding them. It nevertheless still requires special skills and a considerable human effort, as QA testers must manually encode test scripts in the testing framework and must be able to do it in a platform-independent way, as also confirmed by our interviews.

To help QA testers in this difficult task, we propose a new technique for supporting testing of Android apps that has three main capabilities. *First*, it allows testers to interact with an app and both (1) record the actions they perform on the app and (2) specify the expected results of such actions using a new, intuitive mechanism. *Second*, it automatically encodes the recorded actions and specified expected results in the form of a general, platform-independent test script. *Third*, it allows for automatically running the generated test scripts on any platform (*i.e.,* device and operating system), either on a physical device or in an emulator.

In addition, there are several advantages to our approach, compared to the state of the art. One advantage is that our approach implements the record once-run everywhere principle. Testers can record their tests on one platform and ideally rerun them on any other platform. Existing approaches focused on GUI test automation through record/replay [45] tend to generate tests that are brittle and break when run on platforms other than the one on which they were recorded, as confirmed by our empirical evaluation (Section 6). A second advantage of our approach is that it supports the creation of oracles, and it does it in an intuitive way, whereas most existing approaches have very limited support for this aspect [25–27, 45]. In general, our approach can be used with very limited training, as it does not require any special skill or knowledge. A third advantage is that, because of the way our approach encodes test cases, the generated tests tend to be robust in the face of (some) changes in the user interface of the app (and are unaffected by changes that do not modify the user interface). The test cases generated by our approach can therefore also be used for regression testing. From

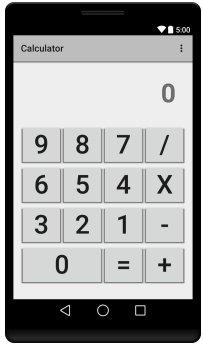
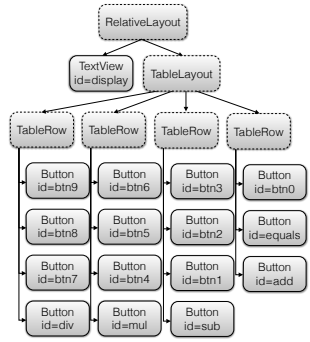

```
1  public void testDivideByZero() {
2      onView(withId(R.id.btn5)).perform(click());
3      onView(withId(R.id.display))
              .check(matches(withText("5")));
4      onView(withId(R.id.divide)).perform(click());
5      onView(withId(R.id.display))
              .check(matches(withText("/")));
6      onView(withId(R.id.btn0)).perform(click());
7      onView(withId(R.id.display))
              .check(matches(withText("0")));
8      onView(withId(R.id.equals))
              .check(matches(isClickable()));
9      onView(withId(R.id.equals)).perform(click());
10     onView(withId(R.id.display))
              .check(matches(withText("ERROR")));
11 }
```

Figure 1: CALCULATOR app.   Figure 2: CALCULATOR UI hierarchy.

Figure 3: Divide-by-zero test case example for CALCULATOR app.

a more practical standpoint, a further advantage of our approach is that it generates test cases in a standard format—the one used in the Espresso framework, in our current implementation. The generated test cases can therefore be run as standalone tests. A final, also practical and advantage of our approach is that it is minimally intrusive. Because it leverages accessibility mechanisms already present on the Android platform [10], our approach does not need to instrument the *apps under test* (*AUTs*). To use the approach, testers only have to install an app on the device on which they want to record their tests, enable the accessibility framework for it, and start recording.

Our technique offers these advantages while handling several practical challenges specific to the Android framework. First, the information required for replay is not directly available from accessibility events, and our technique needs to reconstruct it. This is particularly challenging in our context, in which BARISTA runs in a separate sandbox than the AUT. This challenge is also a distinguishing factor with respect to related work [40] from a related domain (web app) that instead relies on a debugging interface and has direct access to the AUT. Second, our technique must process events in a timely fashion, in the face of a constantly evolving user interface. To address this challenge, our technique efficiently caches the GUI hierarchy and performs operations on its local cache.

To evaluate the practical usefulness and applicability of our technique, we implemented it in a prototype tool, called BARISTA, that encodes user recorded test cases and oracles as Espresso tests. We then performed a comparative user study in which 15 participants used both BARISTA and TESTDROID RECORDER (TR) [24, 45], another state-of-the-art test recording tool, to generate tests for a set of 15 Android apps. The results of this initial study are promising. In particular, they show that BARISTA (1) can faithfully record and encode most user defined test cases, whereas TR fails to do so in many cases, (2) can generate test cases that run on multiple platforms, unlike TR, and (3) provides better support for oracle generation than TR. In more general terms, our evaluation shows that BARISTA has the potential to improve considerably the way test cases for Android apps are generated and run, which can in turn result in an overall improvement of the Android QA process.

In summary, the main contributions of this paper are:

- A technique for easily recording, encoding in a standard format, and executing in a platform independent manner test cases for Android apps.
- An implementation of the technique, BARISTA, that generates Espresso test cases and is freely available for download at http://barista.us-west-2.elasticbeanstalk.com.
- A user study, performed on a set of Android apps, that shows initial yet clear evidence of the practical usefulness and applicability of our technique, together with the improvements it provides over the state of the art.

## 2. MOTIVATING EXAMPLE

In this section, we introduce a motivating example that we use to present underlying concepts behind GUI[1] testing and motivate our work. Figure 1 shows a CALCULATOR app (com.calculator is its package name) that performs arithmetic operations on integers. Its UI consists of (1) a set of buttons for entering digits and operators and computing results and (2) a display area for showing the operands and operators entered by the user and the results of the computation. The calculator handles error conditions, such as division by zero, by displaying text ERROR.

Figure 2 shows the UI hierarchy for our example. The UI components on the screen are the display area (TextView) and a set of Button elements. Each UI element in the app has an identifier, shown with the id property. The arrangement of these UI elements on the screen is managed by the ViewGroup elements in which they are contained, namely, RelativeLayout, TableLayout, and TableRow, which are outlined with dotted lines in the figure.

Manual testing of this app for division-by-zero errors could be performed by simply clicking buttons 5, /, 0, and =, and then checking that the displays shows text ERROR. Although this task is very straightforward, it must be repeated at every test cycle.

As an alternative to the manual approach, Figure 3 shows an equivalent automated UI test. The test first inputs digit 5 into the calculator by clicking the corresponding button, identified by the resource ID R.id.btn5 (line 2). Once this operation is completed, it checks whether the text in the display area, identified by R.id.display (line 3), matches the provided input. The test then performs analogous operations for operator / and digit 0 (lines 4–7). After entering operands and operator, the test checks that button = is clickable (line 8), clicks it (line 9), and then checks that the app suitably shows text ERROR in the display area (line 10).

Although having an automated and easy to rerun test such as this one is a considerable improvement over the manual testing described earlier, creating these tests is still a painful process. First of all, these tests are tedious to write. Furthermore, while writing these tests developers must specify the correct identifier for each interaction and assertion defined. To do so, developer must either analyze layout files in the source code of the AUT or use the UIAUTOMATORVIEWER tool [13] provided as part of the Android SDK. Either way, developers need to alternate between the two tasks of finding identifiers and writing test cases, which further contributes to making test creation a time consuming, tedious, and error prone activity. Our technique, which we describe in the next section, aims to improve the state of the art by bringing together the simplicity of manual testing with the advantages provided by automated testing.

---

[1]The term UI is typically preferred in the context of Android, so we use UI hereafter to refer to the GUI of an Android app.

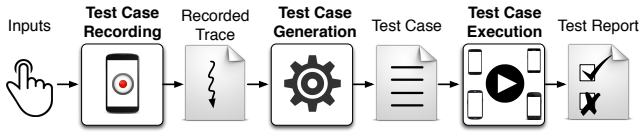

Figure 4: High-level overview of the technique.

## 3. TECHNIQUE

In this section, we present our technique for recording, generating, and executing test cases for Android apps. Figure 4 provides a high-level overview of our technique, which consists of three main phases. In the *test case recording phase*, the user interacts with the AUT with the goal of testing its functionality. Our technique records user interactions and offers a convenient interface to define assertion-based oracles. When the user signals the end of the recording phase, the technique enters its *test case generation phase*, which translates recorded interactions and oracles into test cases that are (as much as possible) device independent. Finally, in the *test case execution phase*, our technique executes the generated test cases on multiple devices and summarizes the test results in a report. In the remainder of this section, we describe these three phases in detail and demonstrate them on our motivating example.

### 3.1 Test Case Recording

In the test case recording phase, the user records test cases by exercising the functionality of an app. This phase receives the package name of the AUT as input. To record the divide by zero test case of Section 2, for instance, the user would indicate `com.calculator` as input of this phase.

Based on the package name provided, the technique launches the app's main activity [17] and, at the same time, creates a *menu*. The menu is displayed as a floating menu above the AUT and is movable, so that it does not interfere with the user interaction with the app. The elements in the menu allow the user to (1) define assertion-based oracles, (2) use system buttons (*i.e.,* back and home buttons), and (3) stop the recording.

As soon as the app is launched, and the menu is visible to the user, a second component starts operating: the *recorder*. This component, which is the core component of the test case recording phase, is used to (1) access the UI displayed by the AUT, (2) process user interactions, and (3) assist the oracle definition process. The recorder leverages the accessibility functionality provided by the Android platform to register for certain kinds of events and be notified when such events occur. The recorder uses these accessibility capabilities to listen to two categories of events: events that describe a change in the UI and events that are fired as consequence of user interactions. Events in the former category are used to create a reference that uniquely identifies an element in the app's UI. We call this reference the *selector* of the element. Events in the latter category, instead, are logged in the recorded trace. Specifically, the recorder (1) stores the type of interaction, (2) identifies the UI element affected by the interaction and defines a selector for it, and (3) collects relevant properties of the interaction. The recorder processes oracles in a similar fashion: it (1) stores the type of oracle, (2) identifies the UI element associated with the oracle and defines a selector for it, and (3) saves the details of the oracle (*e.g.,* an expected value for a field). These interactions and user defined oracles are logged by the recorder in a *recorded trace* in the form of *actions*. When the user stops the recorder, our technique passes the content of the recorded trace to the test case generation phase.

In the rest of this section, we discuss the information collected in the recorded trace, describe how the recorder defines selectors,

```
trace-def         ::= trace main-activity actions
main-activity     ::= string
actions           ::= action | action, actions
action            ::= interaction-def | assertion-def | key-def
interaction-def   ::= interaction i-type selector timestamp i-props
i-type            ::= click | long click | type | select | scroll
selector          ::= resource-id | xpath | properties-based
resource-id       ::= string
xpath             ::= string
properties-based  ::= element-class element-text
element-class     ::= string
element-text      ::= string
timestamp         ::= number
i-props           ::= | exprs
assertion-def     ::= assertion a-type selector timestamp a-props
a-type            ::= checked | clickable | displayed | enabled | focus
                    | focusable | text | child | parent | sibling
a-props           ::= | selector | exprs
key-def           ::= key key-type timestamp
key-type          ::= action | close
exprs             ::= expr | expr, exprs
expr              ::= bool | number | string
```

Figure 5: Abstract syntax of the recorded trace.

present what type of interactions are recognized by our technique, and finally describe the oracle creation process.

#### 3.1.1 Recorded Trace

Figure 5 shows the abstract syntax for a recorded trace. The beginning of the trace is defined by the $trace\text{-}def$ production rule, which indicates that a trace consists of the name of the main activity followed by a list of actions. The types of actions logged into the recorded trace is indicated by the $action$ production rule.

#### 3.1.2 Selectors

Our technique creates a selector for all interactions and oracles, which is used to accurately identify the UI element associated with these actions and is independent from the screen size of the device used in this phase. The technique defines and uses three types of selectors: (1) the *resource ID selector* ($resource\text{-}id$ in Figure 5), (2) the *XPath selector* ($xpath$), and (3) the *property-based selector* ($property\text{-}based$). The resource ID selector corresponds to the Android resource ID that is associated to a UI element [11]; the XPath [47] selector identifies an element based on its position in the UI tree (as the UI tree can be mapped to an XML document); and the property-based selector identifies an element based on two properties: the class of the element ($element\text{-}class$) and the text displayed by the element, if any ($element\text{-}text$).

Our technique does not use the Android resource ID as its only type of selector for two reasons. First, the Android framework does not require a developer to specify the resource ID value for a UI element. In fact, while creating a layout file of an app, it is possible to omit the resource ID of UI elements declared in it. Second, the framework cannot enforce uniqueness of IDs in the UI tree.

In addition, our technique does not use an element's screen coordinates as a selector because the Android ecosystem is too fragmented in terms of screen sizes; the screen coordinates of a UI element on a given device can considerably differ from the coordinates of the same element on a different device.

The recorder aims to identify the most suitable type of selector for every interaction and oracle processed by leveraging the accessibility functionality of the Android platform. It does so by analyzing the accessibility tree representing the UI displayed on the device. Each node in the tree represents an element in the UI and is characterized by two properties of interest: resource ID (if defined) and class of the node (*i.e.,* the class of the UI element represented by the node). The recorder navigates the accessibility tree to track uniqueness of resource IDs. More specifically, the recorder creates a map, which we call the *resource ID map*, where keys are resource IDs of nodes in the tree, and the value associated to each key is the number of nodes having a specific resource ID. To populate the resource ID map, the recorder listens for two types of accessibility event: `TYPE_WINDOW_STATE_CHANGED`, which is fired when the foreground window of the device changes (*e.g.,* a new `Activity` is launched, a `PopupWindow` appears, a `Dialog` is shown), and `TYPE_WINDOW_CONTENT_CHANGED`, which is fired when the content inside a window changes (*i.e.,* when there is a partial change in the UI). Each time one of these two types of event is fired by the Android system, the recorder populates the resource ID map with the information contained in the accessibility tree through a breadth-first traversal.

The information stored in the resource ID map is then used every time an interaction occurs or an oracle is defined by the user. More precisely, when the recorder processes these types of actions, it considers the accessibility node associated with the action. The recorder checks whether the node has a resource ID and, if it does, checks for its uniqueness using the resource ID map. In case the resource ID is unique, the recorder creates a selector of type resource ID for that action. The test case in Figure 3 uses this type of selector for all of its actions. For example, the action at line 2 uses `R.id.btn5` as selector for the button representing number `5`.

If the node associated to an action does not have a resource ID or the ID is not unique, the recorder generates a selector of type XPath. The XPath selector is a path expression that identifies a specific node in the tree. To illustrate, the XPath selector for the button representing number `5` in the test case in Figure 3 would be `/RelativeLayout/TableLayout[2]/TableRow[2]/Button[2]`.

When the window containing the element affected by an interaction transitions to the inactive state immediately after the interaction is performed (*e.g.,* selection on a `ListPreference` dialog), the accessibility framework does not provide the reference to the node in the accessibility tree affected by the interaction. In this case, the recorder cannot define a resource ID or XPath selector and uses a property-based selector instead. The property-based selector leverages the information stored in the accessibility event representing the interaction (see Section 3.1.3 for more details on events). This type of selector identifies an element in the UI using the class of the element and the text displayed by the element (if any). We selected these two properties because they will not change across devices with different screen properties. For the test case in Figure 3, the property-based selector for the button representing number `5` would have `Button` as class and `5` as text.

In general, whenever possible, we favor the use of resource IDs over XPath and property-based selectors because developers explicitly defined such IDs in the source code. Using them should favor readability and understandability of the generated tests.

### 3.1.3 Interactions

The recorder recognizes user interactions by analyzing accessibility events, which are created by the Android platform as a result of such interactions. These events have a set of properties that describe the characteristics of the interactions. For the sake of space, we illustrate how the recorder processes two types of events. Other events are handled following similar mechanisms.

**Click**: Our technique detects when a user clicks on a UI element by listening to accessibility events of type `TYPE_VIEW_CLICKED`. The recorder encodes an event of this type as an entry in the recorded trace ($interaction\text{-}def$ in Figure 5). More specifically, it labels the entry as of type *click* ($i\text{-}type$), identifies the interaction selector ($selector$) as discussed in Section 3.1.2, and saves the action timestamp ($timestamp$). In the case of the example presented in Section 2, the recorder would receive four events of this type: for the clicks on buttons `5`, `/`, `0`, and `=`.

**Type**: Our technique recognizes when a user types text into an app by processing accessibility events of type `TYPE_VIEW_TEXT_CHANGED`. However, naively recording events from this class would translate into having a recorded trace that is not accurate. In fact, the accessibility framework generates an event of this type even when the text is "typed" programmatically as the result of a computation. Our technique addresses this situation by using a finite-state machine (FSM). The FSM leverages the fact that a user-typed text is always followed by an event of type `TYPE_WINDOW_CONTENT_CHANGED`. If the FSM observes this sequence, it enters its accept state and records the event. Otherwise, the FSM enters its reject state and ignores the event. Upon accepting an event of this type, the recorder encodes the event as an entry in the recorded trace ($interaction\text{-}def$), labels the entry as of class *type* ($i\text{-}type$), identifies the interaction selector ($selector$), saves the action timestamp ($timestamp$), and adds the text typed by the user to the properties of the entry ($i\text{-}props$). The Android system fires this type of event every time a user changes the text contained in a text editable element. For this reason, text incrementally typed by a user generates a sequence of events. This sequence of events is processed in the test case generation phase to minimize the size of generated test cases (see Section 3.2). After typing text, a user can click the input method action key (placed at the bottom-right corner of the on-screen keyboard) to trigger developer defined actions. The Android system does not generate accessibility events for this type of interaction. To address this problem, our technique defines a on-screen keyboard that can be used by the tester as a regular keyboard but records this type of interaction as well. In response to this event, the recorder adds an entry ($key\text{-}def$) to its recorded trace (**action**). Our technique handles in a similar fashion the key that, when clicked, hides the on-screen keyboard (**close**).

### 3.1.4 Oracles

Oracles are an essential part of a test case. Our technique supports definition of assertion-based oracles, where assertions can fulfill one of two purposes: either check the state of a UI element at a specific point of the execution or check the relationship of an element with another element in the UI. In the former case, the assertion checks for the value of a specific property characterizing an element state. In the latter case, the assertion checks the relationship between two elements based on their location in the UI tree. Table 1 reports the properties that can be asserted using our technique and provides a brief description of them. Variations of the properties listed in Table 1 can also be asserted. For instance, our technique can be used to assert that the percentage of visible area of an element is above a user defined threshold. Moreover, the technique can also define assertions that check that a property of an element does not have a certain value.

The menu and the recorder contribute together to the creation of assertions. Figures 6, 7, and 8 show part of the assertion creation process. The user starts the process for defining an assertion by clicking the *assert button* in the menu (the button with the tick symbol in Figure 6). The menu then creates the *assertion pane*, a see-through pane that overlays the device screen entirely (Figure 7). This pane intercepts all user interactions until the end of the assertion definition process and is configured so that the Android system does not generate accessibility events for interactions on the pane. In this way, the interactions performed to create the assertion are not included in the recorded trace. At this point, the user can define assertions following one of two processes: automatic or manual.

Table 1: Assertable properties in the test case recording phase.

| Property | Description |
|---|---|
| CHECKED | The element is checked |
| CLICKABLE | The element can be clicked |
| DISPLAYED | The element is entirely visible to the user |
| ENABLED | The element is enabled |
| FOCUS | The element has focus |
| FOCUSABLE | The element can receive focus |
| TEXT | The element contains a specific text |
| CHILD | Child-parent relationship between two elements in the UI |
| PARENT | Parent-child relationship between two elements in the UI |
| SIBLING | Sibling relationship between two elements in the UI |

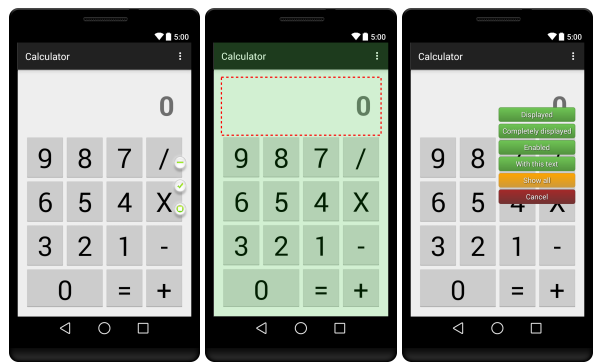

Figure 6: Menu overlay.　　Figure 7: Assertion pane.　　Figure 8: Oracle selection.

**Automatic process**: In this case, the user selects an element in the UI, and our technique automatically adds assertions for each property of the element. The user identifies an element by clicking on the screen and the click is intercepted by the assertion pane, which passes the $x$ and $y$ coordinates of the click location to the recorder. The recorder, upon receiving these coordinates, navigates the accessibility tree to find the node that represents the clicked element (*i.e.,* the foreground node that encloses the coordinates). After identifying the node, the recorder checks the class of the UI element represented by the node and, based on this information, creates assertions to check the value of "relevant" properties of the element. Such relevant properties are defined in a list that we built by analyzing all UI elements in the `android.widget` package. To illustrate, our technique identifies that DISPLAYED, ENABLED, and CLICKABLE are relevant properties for a `Button` element. For each asserted property, the recorder creates an entry in the recorded trace ($assertion\text{-}def$), suitably labels the entry based on the property being checked ($a\text{-}type$), identifies the selector for the assertion ($selector$) as described in Section 3.1.2, and adds the current value of the property to the properties of the entry ($a\text{-}props$).

**Manual process**: In this case, assertions are defined directly by the user. As shown in Table 1, the user can assert properties that affect either a single element or a pair of elements. We illustrate how the technique works when asserting properties that affect one element. (Assertions that affect a pair of elements are defined similarly.) The user selects an element in the UI by long clicking (tap-hold-release) on it. Also in this case, the element is not affected by this action because the assertion pane intercepts the click. In response to the long click, the menu sends the $x$ and $y$ coordinates of the location being pressed to the recorder. The recorder explores the accessibility tree to find the node identified by the location, computes the screen location of the node's vertexes, and sends these coordinates back to the menu. The menu uses the coordinates to highlights the element, as shown in Figure 7.

The user can then change the currently selected element by dragging the finger through the UI elements or accept the currently selected element by raising their finger from the screen. At this point, the recorder identifies the node on the accessibility tree as usual (in case the user changed it), checks the node class, and based on this information sends a list of assertable properties to the menu. The top of the list is populated with properties that are relevant to the node. As shown in Figure 8, these properties are displayed in the proximity of the selected element by the menu. The user can then choose the particular property and the value to be considered in the assertion, and the menu sends the property and the value to the recorder. The recorder creates an entry in the recorded trace ($assertion\text{-}def$), suitably labels the entry based on the selected assertion property ($a\text{-}type$), identifies the selector for the assertion ($selector$), and adds the user defined value for the assertion to the properties of the entry ($a\text{-}props$). Figures 6, 7, and 8 illustrate the manual definition process for the assertion at line 2 in the test case of Figure 1. This assertion checks that, after clicking button `0`, the number `0` appears in the display area (`TextView` element) of the app. After starting the assertion definition process by clicking the button with the tick symbol (Figure 6), the user selects the `TextView` element by long clicking on it (Figure 7) and then selects `With this text` option to define the assertion (Figure 8).

In both cases, after the recorder successfully adds the assertion to its recorded trace, it signals the end of the assertion definition process to the menu. The menu then removes the assertion pane from the screen, and the user can continue to interact with the app.

### 3.2 Test Case Generation

The test case generation phase receives as input the recorded trace and a user-provided flag (*retain-time flag*) that indicates whether the timing of recorded interactions should be preserved. For instance, if a user sets a 30-seconds timer in an alarm clock app and wants to check with an assertion the message displayed when the timer goes off, he or she would set the retain-time flag to true to ensure that the assertion is checked 30 seconds after the timer is started. The test case generation phase produces as output a test case that faithfully reproduces the actions performed by the user during the test case recording phase. In the current version of our technique, the generated test case is an Android UI test case based on the Espresso framework [15]. In the remainder of this section, we illustrate how the technique translates the recorded trace into a test case, discuss the structure of the generated test case, and present the working mechanism of such test case.

The content of a generated test case is divided into two parts: a part that prepares the execution of the test case (*set-up*) and a part the contains the actual steps of the test (*steps*). The two parts are arranged so that the set-up part will execute before the steps part.

The goal of the set-up is to load the starting activity of the test case. This phase retrieves the value of the activity from the recorded trace (see $main\text{-}activity$ in Figure 5) and adds a statement to the set-up section of the test case that loads the activity. This step is necessary in order to align the starting point of the recorded execution with that of the test case.

To generate the steps section of the test case, the technique processes all actions contained in the recorded trace ($actions$) and generates a single-statement line for each one of them. The generated test case thus contains a one-to-one mapping between actions and statements. We believe that this characteristic favors readability and understanding of generated test cases, thus addressing a well-known problem with automatically generated tests. Test case statements that reproduce interactions and oracles are divided into three parts. The first part is used by the test case execution engine to retrieve the UI element affected by the action. Our technique places the selector (*selector*) of the action in this part of the statement. The second part of the statement consists of the action that the test case execution engine performs on the UI element identified by the first part of the statement. The technique encodes this part of the statement with the Espresso API call corresponding to the action being processed (*i-type* or *a-type*). The third part of the statement accounts for parameters involved in the action and is action specific. To create this part, our technique retrieves the properties of the action (*i-props* or *a-props*). For the click of button 5 of the example in Figure 3, the test case generation phase produces `onView(withId(R.id.btn5))` as the first part of the statement and `perform(click())` as its second part. This statement correspond to the statement at line 2 of Figure 3.

The content of the generated test case is affected by the retaintime flag as follows. If the flag is set, our technique places an additional statement between statements representing two subsequent actions. This statement pauses the execution of the test cases (but not the execution of the app being tested) for a duration that is equal to the difference of the timestamps (*timestamp*) associated with the two actions.

### 3.3 Test Case Execution

The test case execution phase takes as input the test case produced by the second phase of the technique, together with a user-provided list of devices on which to run the test case, and performs three main tasks: (1) prepare a device environment for the test case execution, (2) execute the test case, and (3) generate the test report.

The first step installs the AUT and the generated test case on all devices in the user-provided list. Once the execution environment is set up, the technique executes the test case on each device in the user-provided list in parallel. The execution of a test case is supported through our extension of the Espresso framework and works as follows. The test case execution engine loads the starting activity of the test case. From this point forward, the engine synchronizes the execution of the test case's steps with the updates in the UI of the AUT. The engine processes interaction and oracle statements as follows. For both types of actions, it first navigates the UI displayed by the device to find the UI element referenced by the action. If the element is not present, the execution of the test case terminates with an error. If the element is present, the execution engine behaves differently according to whether it is processing an interaction or an oracle statement. In the former case, the execution engine injects a motion event into the app or performs an API call on the UI element being targeted by the interaction. In the case of an oracle statement, the execution engine retrieves all elements in the UI that hold the property expressed by the oracle's assertions and checks whether the element targeted by the oracle is one of these elements. If the element is not present, the test case terminates with a failure. Otherwise, the execution continues. Considering the statement at line 2 of Figure 3, the execution engine would first retrieve the `Button` element represented by the `R.id.btn5` resource ID. It would then inject into the app a motion event that targets the element and produces a click action and wait for the number 5 to be shown in the display area of the app. It would finally execute the statement at line 3.

At the end of the execution, the technique generates a test case execution report that contains: (1) the outcome of the test case on each device, (2) the test case execution time, and (3) debug information if an error or failure occurred during execution.

## 4. IMPLEMENTATION

We implemented our technique in a framework called BARISTA. There are three main modules in the framework: (1) the *recording module*, which implements the aspects of the test case recording phase (Section 3.1); (2) the *generation module*, which generates test cases as presented in the test case generation phase (Section 3.2); and (3) the *execution module*, which executes test cases as described in the test case execution phase (Section 3.3). The recording module is implemented as an Android app and runs on devices that use the platform API level 16 and above. The app does not require root access to the device to operate and does not require the device to be connected to an external computational unit during recording, as the test case recording happens directly and entirely on the device. The generation and execution modules are part of a web service implemented using Java EE [34]. We describe these three components in more detail.

### 4.1 BARISTA App

There are three fundamental components in the BARISTA app: (1) the *menu component*, (2) the *recording component*, and (3) the *input method component*. The three components correspond, respectively, to the menu, recorder, and keyboard presented in Section 3.1. The three components run in distinct processes, which in turn are different from the process in which the AUT is running. This design allows BARISTA to perform its test case recording phase on all apps installed on the device without the need to instrument these apps.

The menu component is implemented as a `Service` [12] and receives messages from other components through a `Broadcast Receiver` [12]. The visual elements of the menu component use the `TYPE_SYSTEM_ALERT` layout parameter, which allow the menu to sit on top of the AUT. The recording component is implemented as an `AccessibilityService` [10] and receives messages from other components through a `BroadcastReceiver`. Finally, the input method component is an `InputMethodService` [14]. When the user ends the recording phase, the app attaches the trace to an HTTP requests and sends it to the BARISTA web service.

### 4.2 BARISTA Web Service

The generation module uses the JavaWriter 2.5 library [42] to create the source code of the generated test cases. BARISTA generates test cases based on the Espresso 1.1 framework [15]. More precisely, BARISTA extends Espresso to provide a larger API that implements the concepts introduced by the technique. The extended API includes the notion of XPath selector (added to the `ViewMatcher` class), a select action for multiple view elements (implemented by extending the `ViewAction` class), and an extended support for the scroll functionality. The BARISTA web service uses the adb server to prepare device environments and execute test cases. Test execution reports are generated using Spoon 1.1 [43].

## 5. LIMITATIONS

As we stated in Section 3.1, our technique leverages the accessibility functionality of the Android platform to detect user interactions. In this way, the technique does not need to run on a "rooted" device, does not need customization of the underlying platform, and does not need to instrument the AUT. However, the accessibility infrastructure does not currently offer support for complex multi-touch gestures (*e.g.,* pinch in and out) and does not fire accessibility events for `WebView` elements. We are currently investigating ways to address these limitations.

Our technique binds interactions and oracles with UI elements. Certain Android apps, however, rely on bitmapped (rather than UI)

elements. Hence, the technique cannot currently handle such apps. Luckily, the vast majority of these apps are games, whereas other types of app tend to rely exclusively on standard UI elements.

When using a property-based selector to identify entities in the app (see Section 3.1.2), entities that belong to the same class and display the same text would have the same selector and would thus be indistinguishable. Although this could be problematic, this type of selector is used only when the resource ID and XPath selectors cannot be used, which is not a common situation. Additionally, this was not a problem in our evaluation.

Finally, BARISTA does not support all aspects and states of the Android activity lifecycle. It nevertheless supports most of them (*e.g.,* pausing, stopping, resuming, restarting), and is able to suitably record transition among states, encode them within test cases (by means of an extension to Espresso that we developed), and suitably replay them.

# 6. EMPIRICAL EVALUATION

To assess the expressiveness, efficiency, and ultimately usefulness of our approach, we used BARISTA to perform a user study involving 15 human subjects and 15 real-world Android apps. Because defining oracles is a fundamental part of generating test cases and of our approach, to perform an apple-to-apple comparison we used as a baseline for our evaluation TESTDROID RECORDER (TR) [24, 45], a popular and freely available state-of-the-art tool that allows users to record tests and define oracles intuitively during recording. We therefore did not consider pure record/replay tools with no oracle definition capabilities, such as RERAN [9], VALERA [22], and MOSAIC [52]. We considered including ACRT [27] in our study, as it can record tests in Robotium [51] format. Unfortunately, however, ACRT does not work with recent Android versions, so using it would have required us to backport our benchmark applications to an earlier Android version.

In our empirical evaluation, we investigated the following research questions:

**RQ1:** Can BARISTA record user defined test cases? If so, how does it compare to TR?

**RQ2:** Is the test case recording process with BARISTA more efficient than the one with TR?

**RQ3:** Does BARISTA's encoding preserve the functionality of the test cases? How does BARISTA compare to TR in this respect?

**RQ4:** Can the test cases generated by BARISTA run on different devices? How platform independent are they with respect to test cases generated by TR?

In the remainder of this section, we first describe the benchmarks used in the evaluation. We then present the user study, discuss evaluation results, and conclude illustrating anecdotal evidence of BARISTA's usefulness using feedback from developers that used it.

## 6.1 Experimental Benchmarks

For our empirical evaluation, we used a set of real-world Android apps. Specifically, we selected 15 free and open-source apps from the F-Droid catalog [8]. Our choice of apps is based on three parameters: (1) popularity, (2) diversity, and (3) self-containment. As a popularity measure, we used the number of installations for an app according to the Google Play store [16]. We selected apps from different categories to have a diverse corpus of benchmarks and prioritized apps for which we did not have to build extensive stubs (*e.g.,* apps that do not rely on a hard to-replicate backend database). Table 2 shows the lists of apps we used. For each app, the table shows its ID (*ID*), name (*Name*), category (*Category*), the range of its installations (*Installations*), and the number of lines of code (*LOC*). The most popular app is BARCODE SCANNER, whose number of installations ranges between 100 and 500 million.

Table 2: Description of our benchmark apps.

| ID | Name | Category | Installations (#K) | LOC (#K) |
|---|---|---|---|---|
| A1 | DAILY MONEY | Finance | 500 - 1000 | 10.7 |
| A2 | ALARM KLOCK | Tools | 500 - 1000 | 6.1 |
| A3 | QUICKDIC | Books | 1000 - 5000 | 289.7 |
| A4 | SIMPLE C25K | Health | 50 - 100 | 1.5 |
| A5 | COMICS READER | Comics | 100 - 500 | 8.4 |
| A6 | CONNECTBOT | Communication | 1000 - 5000 | 24.3 |
| A7 | WEATHER NOTIFICATION | Weather | 100 - 500 | 13.2 |
| A8 | BARCODE SCANNER | Shopping | 100000 - 500000 | 47.9 |
| A9 | MICDROID | Media | 1000 - 5000 | 5.6 |
| A10 | EP MOBILE | Medical | 50 - 100 | 31.4 |
| A11 | BeeCount | Productivity | 10 - 50 | 16.2 |
| A12 | BODHI TIMER | Lifestyle | 10 - 50 | 10.5 |
| A13 | ANDFHEM | Personalization | 10 - 50 | 60.3 |
| A14 | XMP MOD PLAYER | Music & Audio | 10 - 50 | 58.7 |
| A15 | WORLD CLOCK | Travel & Local | 50 - 100 | 31.4 |

## 6.2 User Study

For our experimentation, we recruited 15 graduate students from three institutions. We asked the participants to perform three tasks: (1) write natural language test cases (NLTCs), (2) record NLTCs using TR, and (3) record NLTCs using BARISTA. All participants started from the first task. Eight of them performed the second task before the third and the remaining seven did the opposite. Before performing the user study, we conducted a two-hour tools demonstration session to familiarize the participants with the two tools. We did not inform the subjects of which tool was ours and which one was the baseline (but they obviously could have discovered this by searching the name of the tools).

In the first task we provided the participants with three benchmark apps, so that each app was assigned to three different users. We asked the participants to explore the apps' functionality and then define five NLTCs for each app assigned to them. NLTCs were written purely in natural language, without the use of any framework and without even following any particular structure. After they all completed the first task, we manually analyzed the NLTCs for possible duplicates and checked with the participants in case of ambiguities. Table 3 shows the properties of the NLTCs we collected. For each app, the table shows the number of distinct NLTCs (*NLTCs*), average number of interactions per test case (*I*), and average number of assertions per test case (*A*). The total number of distinct NLTCs is 215. All NLTCs have at least one assertion. A1 is the app having the NLTC with the highest number of interactions (27), while A11 is the app with the NLTC having the highest number of assertions (10). All NLTCs are expected to pass.

In the second and third tasks, we asked the participants to record NLTCs using TR and BARISTA, respectively. For each task, each participant was provided with a set of NLTCs written for three apps. The set of NLTCs for the second task was different from the set for the third task. We also decided not to give participants NLTCs they wrote, so as to mimic a scenario in which the test specifications are provided by a requirements engineer and the testing is performed by a QA tester. For each of the two tasks, we asked the users to reproduce the steps of the NLTCs as faithfully as possible, unless the tool prevented them to do so (*e.g.,* they could skip assertions that the tool was unable to encode).

The experimental setup to perform the second task was structured as follows. We asked users to record NLTCs on a device running Android API level 19. The device was connected to a

MacBook Pro (2.3 GHz i7 processor and 8GB memory) running Eclipse 4.4, with TR installed as a plugin. Users could record a test case by starting the plugin and selecting the app to test, which automatically installed the app on the Android device. After this point, users interacted with the app directly to record a test case. To define an assertion using TR, users might need to specify the Android resource IDs of the element involved in the assertion. We thus made the UIAUTOMATORVIEWER tool [13] available to users, so that they could easily explore an app's UI. For the same purpose, we also provided the participants with the source code of the benchmark apps. We did not impose any timeout to perform the task. To perform the third task, we asked users to record NLTCs using a device running API level 19 with BARISTA installed. Also in this case, we did not impose any timeout.

### 6.3 Results

**RQ1**: To answer the part of RQ1 about BARISTA's expressiveness, we checked the test cases recorded by users using BARISTA against the corresponding NLTCs. The first part of Table 4 (columns below BARISTA header) shows the results of this check. For each app, we report the number of test cases that could be recorded (*C*), the number of test cases that could not be recorded (*NC*), the number of assertions skipped (*AS*), and the number of assertion altered (*AA*). We considered an NLTC as recorded if the generated test case contained all interactions defined in it, and not recorded otherwise. We considered an assertion as skipped if the user did not define it, whereas we considered an assertion as altered if the user defined an assertion with a different meaning from the one in the NLTC. When using BARISTA, participants could record all test cases, skipped 11 assertions (2.2% of the total number of assertions), and did not alter any assertion. The 11 assertions that users could not express with BARISTA do not directly check for properties of UI elements (*e.g.,* an NLTC for A4 states "assert that the alarm rings").

The second part of Table 4 (columns below TR header) helps us to answer the second part of RQ1, which compares BARISTA and TR in terms of expressiveness. This part of Table 4 reports, for TR, the same information that we reported for BARISTA in the first part of the table. In this case, 44 test cases could not be recorded. 36 of those could not be recorded because TR altered the functionality of 10 apps (A2, A3, A5, A6, A7, A9, A10, A11, and A13), which resulted in participants not being able to perform certain interactions. Although the tool is not open source, so we could not verify our hypothesis, we believe that this side effect depends on how TR either installs an app on a device or inspects the Android's UI hierarchy while processing user inputs. In six additional test cases, users stopped recording the test case after skipping the first assertion. Finally, one user forgot to record one test case. Even without considering the last seven test cases, which mostly depend on user errors, BARISTA could record 20.5% more test cases than TR.

As the table also shows, users skipped 108 assertions while using TR. (The assertions skipped while using BARISTA are included in this set.) The reason behind this high number is that TR offers a limited range of assertable properties. For instance, TR does not provide assertions to check whether a UI element is clickable or whether an element is checked. Among the test cases recorded by both tools, BARISTA could express 24.6% more assertions than TR. In the test cases generated by TR, we can also note that 48 assertions (sum of column *AA*) were different from the ones defined in the corresponding NLTCs. An example of such assertion mismatch is given by a NLTC from A1, for which the user recorded "assert button is enabled" instead of "assert button is clickable". We asked the participants involved why they modified these assertions, and they said that it was because they could not find a way to record the original assertion with the tool.

These results provide initial evidence that BARISTA can record user defined test cases and is more expressive than TR.

**RQ2**: To answer RQ2, we compared the amount of time taken by the participants to record test cases using BARISTA with the time they needed when using TR. For each app, Table 4 reports the average time difference (*TD*) between BARISTA and TR, expressed as a percentage. The percentage is computed considering the test cases that were recorded by both tools and in which no assertion was skipped. The amount of time associated with each test cases is calculated from the moment in which the user recorded the first action to the time in which the user terminated the recording process. The recording process using BARISTA was faster than TR for all apps. Overall, BARISTA was 37.18% faster than TR. To provide an idea of the amount of time taken to record test cases with BARISTA, the test case with the highest number of interactions (27 interactions and 3 assertions) was recorded in five minutes, while the test case with the highest number of assertions (9 interactions and 10 assertions) was recorded in two minutes.

We can therefore say, for RQ2, that the test case recording process of BARISTA is more efficient than that of TR.

**RQ3**: To answer the part of RQ3 about BARISTA's correctness, we executed the 215 test cases generated using BARISTA on the device on which they were recorded. We report the execution results in the first part of Table 5 (columns below BARISTA header). For each app, we report the number of test cases executed (*T*), the number of test cases that worked correctly (*W*), the number of test cases that terminated with an error or failure due to a problem in the tool generation or execution phase (*NW*), and the number of test cases that terminated with an error or failure due to a user mistake in the recording process (*M*). We consider a test case as working correctly if it faithfully reproduces the steps in its corresponding NLTC. Across all benchmark apps, 97.2% of the recorded test cases worked correctly, and 12 apps had all test cases working properly. The test case from A5, which is marked as not working, terminated with an error because the file system of the device changed between the time the test case was recorded and the time the test case was executed. The five test cases marked as user mistakes terminated with an assertion failure. In two of these cases, the user asserted the right property but forgot to negate it. In the remaining three test cases, the user asserted the right property but on the wrong UI element. We presented the errors to users and they confirmed their mistakes.

The second part of Table 5 (columns below TR header) lets us answer the second part of RQ3, which compares the correctness of the test cases generated by BARISTA with respect to that of the tests generated by TR. The second part of Table 5 reports the classification of execution results for the 171 test cases generated using TR. We executed these test cases on the same device on which they were recorded. Across all benchmark apps, only 64.9% of the recorded test cases worked correctly, which corresponds to 51.6% of the NLTCs. BARISTA, in comparison, nearly doubles this percentage. The 49 test cases classified as not working could not replicate at least on of the interactions from their corresponding NLTCs. Users made 11 mistakes using TR. In the majority of these cases (6), the user entered the wrong resource ID when recording an assertion. As in the case of BARISTA, we presented the errors to the participants responsible, and they confirmed their mistakes.

Based on these results, we can answer RQ3 as follows: there is evidence that test cases generated by BARISTA work correctly, and that BARISTA can outperform TR in this respect.

**RQ4**: To answer the part of RQ4 on BARISTA's cross-device compatibility, we executed the test cases recorded using BARISTA

Table 3: NLTCs properties.

| ID | NLTCs(#) | I(#) | A(#) |
|---|---|---|---|
| A1 | 15 | 9.33 | 3.40 |
| A2 | 15 | 7.07 | 1.40 |
| A3 | 14 | 6.21 | 1.36 |
| A4 | 14 | 4.36 | 3.14 |
| A5 | 14 | 3.50 | 1.93 |
| A6 | 13 | 8.92 | 1.23 |
| A7 | 14 | 3.29 | 2.79 |
| A8 | 14 | 2.93 | 1.86 |
| A9 | 12 | 4.08 | 1.25 |
| A10 | 15 | 6.47 | 3.00 |
| A11 | 15 | 6.73 | 2.20 |
| A12 | 15 | 3.67 | 1.73 |
| A13 | 15 | 3.93 | 3.13 |
| A14 | 15 | 4.87 | 2.47 |
| A15 | 15 | 4.47 | 3.27 |

Table 4: Test case recording process characteristics.

| ID | BARISTA | | | | TR | | | | TD(%) |
|---|---|---|---|---|---|---|---|---|---|
| | C(#) | NC(#) | AS(#) | AA(#) | C(#) | NC(#) | AS(#) | AA(#) | |
| A1 | 15 | 0 | 2 | 0 | 15 | 0 | 9 | 20 | -32.33 |
| A2 | 15 | 0 | 0 | 0 | 4 | 11 | 0 | 2 | -38.71 |
| A3 | 14 | 0 | 2 | 0 | 9 | 5 | 5 | 1 | -20.77 |
| A4 | 14 | 0 | 3 | 0 | 9 | 5 | 8 | 7 | -37.24 |
| A5 | 14 | 0 | 0 | 0 | 12 | 2 | 2 | 2 | -76.57 |
| A6 | 13 | 0 | 0 | 0 | 6 | 7 | 0 | 1 | -36.10 |
| A7 | 14 | 0 | 0 | 0 | 11 | 3 | 13 | 5 | -11.86 |
| A8 | 14 | 0 | 0 | 0 | 11 | 3 | 5 | 0 | -79.86 |
| A9 | 12 | 0 | 0 | 0 | 11 | 1 | 3 | 3 | -53.79 |
| A10 | 15 | 0 | 0 | 0 | 13 | 2 | 10 | 2 | -3.93 |
| A11 | 15 | 0 | 0 | 0 | 12 | 3 | 10 | 0 | -13.30 |
| A12 | 15 | 0 | 0 | 0 | 15 | 0 | 5 | 0 | -11.03 |
| A13 | 15 | 0 | 0 | 0 | 13 | 2 | 14 | 3 | -62.87 |
| A14 | 15 | 0 | 2 | 0 | 15 | 0 | 7 | 2 | -48.13 |
| A15 | 15 | 0 | 2 | 0 | 15 | 0 | 17 | 0 | -31.19 |

Table 5: Test case execution outcome.

| ID | BARISTA | | | | TR | | | |
|---|---|---|---|---|---|---|---|---|
| | T(#) | W(#) | NW(#) | M(#) | T(#) | W(#) | NW(#) | M(#) |
| A1 | 15 | 15 | 0 | 0 | 15 | 8 | 6 | 1 |
| A2 | 15 | 15 | 0 | 0 | 4 | 3 | 1 | 0 |
| A3 | 14 | 14 | 0 | 0 | 9 | 5 | 4 | 0 |
| A4 | 14 | 12 | 0 | 2 | 9 | 3 | 5 | 1 |
| A5 | 14 | 13 | 1 | 0 | 12 | 10 | 0 | 2 |
| A6 | 13 | 13 | 0 | 0 | 6 | 4 | 2 | 0 |
| A7 | 14 | 11 | 0 | 3 | 11 | 9 | 2 | 0 |
| A8 | 14 | 14 | 0 | 0 | 11 | 8 | 2 | 1 |
| A9 | 12 | 12 | 0 | 0 | 11 | 11 | 0 | 0 |
| A10 | 15 | 15 | 0 | 0 | 13 | 9 | 4 | 0 |
| A11 | 15 | 15 | 0 | 0 | 12 | 8 | 4 | 0 |
| A12 | 15 | 15 | 0 | 0 | 15 | 12 | 3 | 0 |
| A13 | 15 | 15 | 0 | 0 | 13 | 1 | 9 | 3 |
| A14 | 15 | 15 | 0 | 0 | 15 | 9 | 5 | 1 |
| A15 | 15 | 15 | 0 | 0 | 15 | 11 | 2 | 2 |

on seven (physical) devices: LG G FLEX (D1), MOTOROLA MOTO X (D2), HTC ONE M8 (D3), SONY XPERIA Z3 (D4), SAMSUNG GALAXY S5 (D5), NEXUS 5 (D6), and LG G3 (D7). (We acquired these devices in early 2015 with the goal of getting a representative set in terms of hardware and vendors.) We executed all the test cases that did not contain a user mistake, and among those, 206 test cases worked on all devices. Overall, the average compatibility rate across all apps and devices was 99.2%. Two test cases (from A13) did not work on D7 because that device adds additional space at the bottom of a `TableLayout` element. The additional space moves the target element of an action out of the screen, preventing BARISTA from successfully interacting with that element. (The two test cases work on D7 by adding a scroll action to the test cases.) Also, one test case (from A13) did not work on D1, D5, and D7 because these devices display an additional element in a `ListView` component. For this reason, an interaction in the test case selects the previous to last element instead of the last element.

To answer the second part of RQ4, which involves comparing cross-device compatibility of test cases generated using BARISTA with respect to the one generated with TR, we executed, on the seven devices considered, the test cases recorded using TR that did not contain a user mistake. Of those, 108 tests worked on all devices, and the average compatibility rate across all apps and devices was 68.3% (compared to BARISTA's 99.2%). Many of these failing tests also fail on the device on which they were recorded. In addition, TR generated three test cases that did not work on D5: one test (from A9) fails to identify the target element of an interaction based on the $x$ and $y$ coordinates stored in the test case; two tests (from A15) do not work because the index used to select the target element of an interaction is not valid on the device. It is worth noting that, whereas for the three BARISTA-generated tests that are not cross-device compatible, the corresponding TR-generated tests are also not cross-device compatible, the opposite is not true. For the TR-generated tests that are not cross-device compatible, the corresponding BARISTA-generated tests are cross-device compatible.

Based on these results, we can answer RQ4 as follows: test cases generated using BARISTA can run on different devices in a majority of cases, and BARISTA generated a greater number of test cases that are cross-device compatible than TR.

### 6.4 Developers Feedback

We recently publicly released BARISTA and also directly contacted several developers in various companies to introduce our tool and ask them to give us feedback in case they used it. Although this is admittedly anecdotal evidence, we want to report a few excerpts from the feedback we received, which echo some of our claims about BARISTA's usefulness. Some feedback indicates the need for a technique such as BARISTA: *"I have been looking for something like BARISTA to help me get into automation for a while"*. Other feedback supports the results of our empirical evaluation on the efficiency of BARISTA: *"Overall, a very interesting tool! For large-scale production apps, this could save us quite some time by generating some of the tests for us"*. Finally, some other feedback points to aspects of the technique that should be improved and that we plan to address in future work: *"There are a few more assertions I'd like to see. For example, testing the number of items in a `ListView`"*. We are collecting further feedback and will make it available on the BARISTA's website after anonymizing it.

## 7. RELATED WORK

In this section, we review the work that is most closely related to BARISTA. In particular, we discuss the techniques that fall in the intersection between record/replay and GUI test automation.

### 7.1 Desktop Techniques

In the domain of desktop apps, there is a large body of techniques that focus on GUI test automation using a record/replay approach [1, 23, 30, 36, 44]. These techniques differ from each other in the way they identify elements in the GUI and the type of actions they record. JACARETO [23] and POUNDER [36] are event-driven techniques that record low level events such as mouse clicks and keyboard strokes. JRAPTURE [44] identifies a GUI element by combining the identifier of the thread that created the element together with the running count of elements created by the thread.

BARISTA can be related to MARATHONITE [30] and ABBOT [1] as they all work at a higher level of abstraction recording semantic actions and identify elements using unique properties of each element rather than using its coordinates. However, BARISTA differs from the two techniques in the way it record actions. MARATHONITE and ABBOT use dynamic instrumentation that is not possible in Android unless of modifications to the Android software stack [32]. BARISTA does not change the Android software stack and instead records actions using the accessibility infrastructure offered by the framework.

### 7.2 Web Techniques

There is also a rich set of techniques for GUI test automation through record/replay in the web app domain [4, 6, 20, 31, 35, 40, 41]. Some of these techniques focus on the execution of JavaScript [31, 41]. JALANGI [41] is one of such techniques and offers a selective record/replay approach that enables recording and replaying of a user-selected part of the program. DODOM [35], WARR [4], and SELENIUM IDE [40] are techniques that focus on user interactions.

DODOM records user interaction sequences and executes the AUT under the recorded sequences to observes its behavior and detect DOM invariants. WARR modifies the WebKit engine in the Chrome web browser to record interactions and implements a browser interaction driver to replay them. SELENIUM IDE is a Firefox add-on that generates test cases in the SELENIUM format.

BARISTA can be related to SELENIUM IDE in that they both record semantic actions and offer the opportunity to express oracles in the recording process. However, SELENIUM IDE has direct access the AUT while BARISTA can access the AUT only upon receiving certain accessibility events, this difference makes the recording task of our technique more challenging. Furthermore, SELENIUM IDE runs with heightened privileges [39], while our approach does not require higher privileges, which would requires modifications to the Android framework. In addition, test scripts generated by SELENIUM IDE are sensitive to possible delays in the execution of a web app. Section 6 shows that BARISTA does not have this issue. Finally, BARISTA can be related to the work from Grechanik and colleagues [20] as they also leverage accessibility technologies to create GUI testing techniques.

### 7.3 Mobile Techniques

In the domain of mobile apps, there are techniques that focus on GUI test automation through record/replay [24–27] and techniques that focus mainly on the record/replay task [9, 22, 37, 52].

TESTDROID RECORDER [24, 45] is commercial tool, implemented as an Eclipse plugin, that records interactions from a connected device running the AUT. The tool allows users to define oracles using the plugin and generates tests that rely on the ROBOTIUM framework [51]. BARISTA is similar to TR in that they both record interactions at the application layer, however the approach used by TR presents some limitations. First, TR uses identifiers that do not reliably identify elements in the GUI. Second, generated tests rely on `sleep` commands, which make tests slow and unreliable. Third, the tool does not suitably instrument the UI of the AUT to process user inputs leading to missed interactions in recorded test cases.

ACRT [27] is a research tool that, similarly to TR, generates ROBOTIUM tests from user interactions. ACRT approach is based on app instrumentation. ACRT modifies the layout of the AUT to record user interactions and adds a custom gesture to certain element of the GUI to allow the user to definition oracles. The record/replay process is performed by the user using an Eclipse plugin. As in the case of TR, ACRT generates tests that rely on `sleep` commands, which make tests slow and unreliable. In addition, the support for interactions and oracles is limited and the technique does not consider how to uniquely identify elements in the GUI.

SPAG [25] uses SIKULI [50] and ANDROID SCREENCAST [5] to create a system in which the screen of the AUT is redirected to a PC and the user interacts with AUT using the PC. The user records scripts by interacting with the PC and oracles are in the form of screenshots. SPAG−C [26] extends SPAG using image comparison methods to validate recorded oracles. The approach for oracle definition presented in SPAG and SPAG−C is minimally invasive, as it does not modify the AUT. However, expressing oracles for a specific element in the GUI is a practical challenge and the image comparison approach can miss small but significant differences.

MOBIPLAY [37] is a record/replay technique based on remote execution. The technique executes the AUT in a virtual machine running on the server and displays the GUI of the app on a client app running on the mobile phone. The inputs to the AUT are recorded using the client application and can be replayed either on the client or the server component. MOBIPLAY is similar to BARISTA in that inputs to the AUT are collected at the application layer, however MOBIPLAY input collection approach requires modifications in the Android software stack. In addition, MOBIPLAY records inputs based on their screen coordinates, while BARISTA collects them so that they are platform-independent.

RERAN [9] records low level system events by leveraging the Android GETEVENTS utility and generates a replay script for the same device. The low level approach presented by RERAN is effective in recording and replaying complex multi-touch gestures. However, generated scripts are not suitable for replay on different devices because recorded interactions are based on screen coordinates. VALERA [22] redesigns and extends RERAN with a stream-oriented record-and-replay approach. VALERA is capable of recording and replaying: touchscreen events, sensor and network inputs, inter-app communication, and event schedules. MOSAIC [52] extends RERAN to overcome the device fragmentation problem. The technique abstracts low-level events into a platform-agnostic intermediate representation before translating them to a target system. VALERA, and MOSAIC are powerful techniques to record platform-independent multi-touch gestures and stream data (VALERA), however they do not support definition of oracles, which constitute a fundamental aspect of GUI testing.

Finally, related work also contains a great amount of research in UI testing and UI exploration techniques [2, 3, 7, 18, 19, 21, 28, 29, 38, 46, 48, 49]. UIAUTOMATOR [19] leverages the accessibility features of the Android framework to perform automated test case execution. BARISTA is similar to UIAUTOMATOR in that it uses the accessibility framework. However, our technique uses the framework for a different purpose. In fact, BARISTA uses the framework for recording accurate execution traces that will then be used to generate test cases.

## 8. CONCLUSION

In this paper, we presented a new technique for helping testers create (through recording), encode (using a standard format), and run (on multiple platforms) test cases for Android apps. One distinctive feature of our technique is that it allows to add oracles to the tests in a visual and intuitive way. We implemented our technique as a freely available tool called BARISTA. Our preliminary evaluation of BARISTA shows that it can be useful and effective in practice and that it improves the state of the art.

There are a number of possible directions for future work. As far as the empirical work is concerned, we will extend our current evaluation by (1) performing an additional user study with a large number of experienced Android developers and (2) running the generated test cases also on different versions of the Android OS.

We will also extend our technique in several ways. First, we will add to BARISTA the ability to factor out repetitive action sequences, such as app initialization, and allow testers to load these sequences instead of having to repeat them for every test. Second, we will investigate how to add sandboxing capabilities to BARISTA, so that it can generate tests that are resilient to changes in the environment. Third, based on feedback from developers, we will extend the set of assertable properties that testers can use when defining oracles. Fourth, we will investigate the use of fuzzing for generating extra tests in the proximity of those recorded, possibly driven by specific coverage goals. Fifth, we will study ways to help developers fix broken test cases during evolution (*e.g.,* by performing differential analysis of the app's UI). Finally, we will consider the use of our technique to help failure diagnosis; a customized version of BARISTA could be provided to users to let them generate bug reports that allow developers to reproduce an observed failure.